\newif\ifarxiv
\arxivtrue %

\documentclass[letterpaper, 10 pt, conference]{ieeeconf}  %
\ifarxiv
    \pdfoutput=1
    \usepackage{hologo}
    \usepackage{listings}
\else
\fi

\IEEEoverridecommandlockouts

\newcommand{\withcoverpage}{1}

\newcommand{\titlestring}{A Prototypical Expert-Driven Approach Towards Capability-Based Monitoring of Automated Driving Systems}
\newcommand{\authorstring}{Richard Schubert, Cedrik Kaufmann, Marcus Nolte, and Markus Maurer}
\newcommand{\authorstringaff}{Richard Schubert$^{1}$, Cedrik Kaufmann$^{1}$, Marcus Nolte$^{1}$, and Markus Maurer$^{1}$}
\newcommand{\yearstring}{2024}

\title{\LARGE
        \bf\titlestring
        $^{*}$%
        \vspace{-0.5em}
}

\author{\authorstringaff%
        \thanks{$^{*}$\scriptsize%
        This research is accomplished within the projects
	\qemph{UNICARagil} (FKZ 16EMO0285) and
	\qemph{AUTOtechagil} (FKZ 01IS22088R).
	We acknowledge the financial support for both projects by
	the Federal Ministry of Education and Research of Germany (BMBF).}%
        \thanks{$^{1}$\scriptsize%
        Institute for Control Engineering, TU Braunschweig, Germany \linebreak
        {\tt \{richard.schubert, c.kaufmann, marcus.nolte, markus.maurer\}@tu-braunschweig.de}}%
}

\newcommand{\conferencestring}{27th IEEE International Conference on Intelligent Transportation Systems}
\newcommand{\addressstring}{Edmonton, Canada}

\usepackage{verbatim}
\usepackage{tabularx,calc}
\usepackage{color}
\usepackage{tubscolors}
\usepackage{adjustbox}
\usepackage{hyperref}
\usepackage{ifthen}
\usepackage{lipsum}
\usepackage{xstring}
\usepackage{amssymb} 	%
\usepackage{amsmath}
\usepackage{siunitx}
\usepackage{graphicx}
\usepackage{bm} 		%
\usepackage{booktabs} 	%
\usepackage{multirow}   %
\usepackage{booktabs} 	%
\usepackage{colortbl}
\usepackage{microtype}
\usepackage{here}
\usepackage{tabularx}
\usepackage{arydshln}
\usepackage{titlesec}
\usepackage{array}
\usepackage{siunitx}
\usepackage{listings}
\usepackage{adjustbox}
\usepackage{titlesec}
\usepackage[bibstyle=ieee,
			citestyle=numeric-comp,
			backend=bibtex,
			minnames=1,
            maxnames=3,
			maxcitenames=1,
			maxbibnames=4,
			doi=false,
			isbn=false,
			url=false,
			natbib=true]
			{biblatex}

\DefineBibliographyStrings{english}{andothers={et\addabbrvspace al\adddot}}

\DefineBibliographyStrings{english}{phdthesis = {PhD Thesis,}}

\AtEveryBibitem{\clearlist{language}}

\newcommand{\qemph}[1]{\emph{#1}}

\renewcommand{\vec}[1]{\mathbf{#1}}

\renewcommand{\dots}{\, ...\,}

\newcommand{\nidx}{i}
\newcommand{\pidx}{j}

\newcommand{\Pri}[2]{P(#1\,|\,#2)}

\newcommand{\Q}{\mathcal{Q}}
\newcommand{\Qi}{Q_{\nidx}}
\newcommand{\Qj}{\mathbf{Q_{\pidx}}}
\newcommand{\Qjo}{Q_{\pidx_1}}
\newcommand{\Qjn}{Q_{\pidx_n}}

\newcommand{\meas}{m} %

\newcommand{\miout}{\meas_\nidx} %
\newcommand{\mjoout}{\meas_{\pidx_1}} %
\newcommand{\mjnout}{\meas_{\pidx_N}} %

\newcommand{\srule}{R} %
\newcommand{\srules}{\mathcal{R}} %
\newcommand{\srulesv}{\mathcal{R}_v} %

\newcommand{\vpi}{\vec{B}_{\nidx}}
\newcommand{\cbehnoidx}{b}
\newcommand{\cbeh}{\cbehnoidx_{\nidx}}

\newcommand{\statev}{S_v}
\newcommand{\statevi}{\statev^{(\nidx)}}

\newcommand{\statevo}{\statev^{(\pidx_1)}}
\newcommand{\statevn}{\statev^{(\pidx_N)}}
\newcommand{\ant}{A}
\newcommand{\con}{C}

\newcommand{\mustatevnn}{\mu_{\pidx_n}^{\statev}}%

\newcommand{\mustateann}{\mu_{\pidx_n}^{\ant_n(\srule)}}
\newcommand{\muobs}{\mu_{i,\mathrm{obs}}^{\statev}}

\newcommand{\good}{\text{``$\mathrm{good}$''}}
\newcommand{\pgood}{\text{``$\mathrm{probably}$ $\mathrm{good}$''}}
\newcommand{\pdgood}{\text{``$\mathrm{prob.}$ $\mathrm{good}$''}}
\newcommand{\pbad}{\text{``$\mathrm{probably}$ $\mathrm{bad}$''}}

\newcommand{\bad}{\text{``$\mathrm{bad}$''}}

\newcommand{\s}{\,\mathrm{s}}
\newcommand{\perc}{\,\%}

\newcommand{\tkmh}{\,\mathrm{km/h}}

\newcommand{\unicar}{UNICAR\emph{agil}}
\newcommand{\autotech}{AUTOtech.\emph{agil}}

\definecolor{accessblue}{cmyk}{1,0.6,0.0,0}
\definecolor{greycolor}{cmyk}{0,0,0,1}

\newcounter{requirement}

\renewcommand{\normalsize}{\fontsize{10}{11.3}\selectfont} %

\newcommand{\titlebaselineskip}{0.25\baselineskip}
\titlespacing*{\section}{-0.75em}{\dimexpr\titlebaselineskip}{\titlebaselineskip}

\newcommand{\eqskipInPt}{5pt}
\AtBeginDocument{%
  \abovedisplayskip=\eqskipInPt
  \abovedisplayshortskip=\eqskipInPt
  \belowdisplayskip=\eqskipInPt
  \belowdisplayshortskip=\eqskipInPt
}

\titleformat{\paragraph}[runin]{\normalfont\normalsize\itshape}{\theparagraph}{0.5em}{\hspace{0.5em}}[:]
\titlespacing*{\paragraph}{0pt}{1ex plus 0.5ex minus 0.2ex}{0.5em}
\renewcommand{\theparagraph}{\alph{paragraph})}

\renewenvironment{quote}
  {\vspace{0.15em}\list{}{\rightmargin=10pt \leftmargin=10pt}%
   \item\relax}
  {\endlist\vspace{0.15em}}

\ifthenelse{\withcoverpage=1}{
        \pdfoutput=1
\usepackage{hologo}
\usepackage{listings}
\lstset{breaklines,basicstyle=\small,columns=fullflexible,basicstyle=\ttfamily,language={[plain]TeX}}
}{}

\begin{document}

\ifthenelse{\withcoverpage=1}{
        \newif\ifpublished
\publishedtrue

\ifpublished
    \twocolumn[
    \begin{@twocolumnfalse}

        \Huge {IEEE Copyright Notice}

        \vspace{0.25cm}
        
        \large {\copyright\ \yearstring\ IEEE. Personal use of this material is permitted. Permission from IEEE must be obtained for all other uses, in any current or future media, including reprinting/republishing this material for advertising or promotional purposes, creating new collective works, for resale or redistribution to servers or lists, or reuse of any copyrighted component of this work in other works.}

        \vspace{1.25cm}
        
        {\Large Published in \emph{\conferencestring}}

        \vspace{0.5cm}

        \textit{Cite as:}
        \vspace{0.2cm}

        \authorstring,
        ``\titlestring,''
        in \emph{\conferencestring},
        \addressstring, \yearstring.

    \end{@twocolumnfalse}
]
\else
    \twocolumn[
        \begin{@twocolumnfalse}
            \textit{Cite as:}
            \vspace{0.2cm}

            \authorstring,
            ``\titlestring,''
            {submitted for publication}.

            \vspace{0.5cm}

            \textit{BibTeX:} \vspace{0.2cm}
        
            \texttt{%
            @inproceedings\{schubert\_conformal\_\yearstring,\\
            author=\{\authorstring\},\\
            title=\{\titlestring\},\\
            year=\{\yearstring\},\\
            publisher=\{submitted for publication\}\\
            \}
            }

        \end{@twocolumnfalse}
    ]
\fi
}{}

\maketitle
\thispagestyle{empty}
\pagestyle{empty}

\vspace{-1.0em}

\begin{abstract}
    Supervising the safe operation of automated vehicles is a key requirement
in order to unleash their full potential in future transportation systems.
In particular,
previous publications have argued that SAE Level 4 vehicles should be aware of their \qemph{capabilities}
at runtime to make appropriate behavioral decisions.
In this paper, we present a framework that enables the implementation of an online capability monitor.
We derive a graphical system model that captures the relationships between
the quality of system elements across different architectural views.
In an expert-driven approach,
we parameterize \qemph{Bayesian Networks} based on this structure
using \qemph{Fuzzy Logic}. %
Using the online monitor, we infer the quality of the system's capabilities
based on technical measurements acquired at runtime.
\end{abstract}

\begin{figure}[H]
    \centering
    \includegraphics[width=0.99\columnwidth, height=3.75cm]{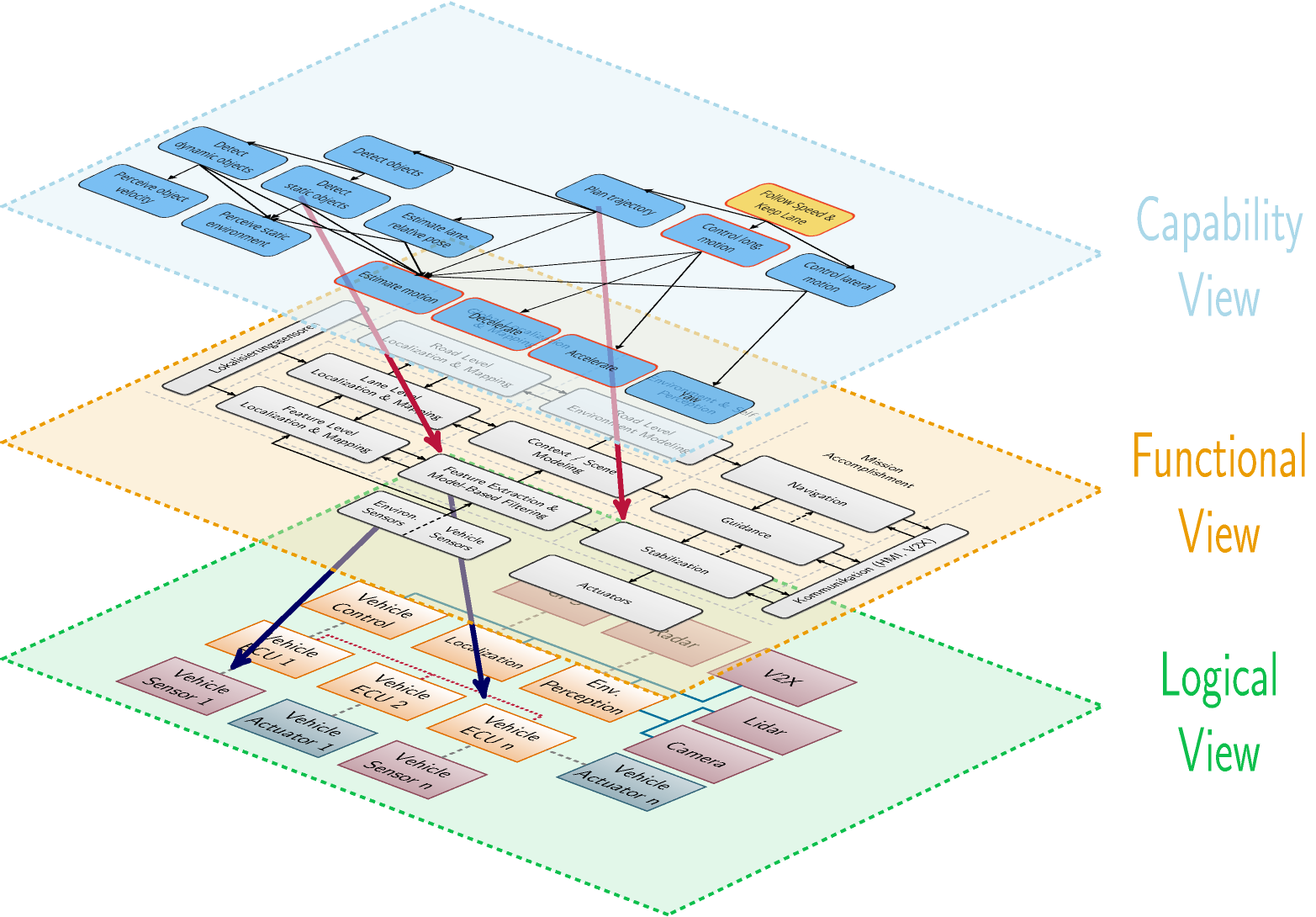}
    \vspace{-0.5em}
    \caption{%
        Architectural viewpoints, adapted from \parencite{bagschik_systems_2018}.
        Arrows indicate correspondences between viewpoints.%
    }
    \label{fig:arch_viewpoints}
    \vspace{-0.5em}
\end{figure}
\section{Introduction} \label{sec:intro}
Ensuring the safe operation of automated vehicles in urban areas is a central research focus.
When aiming to operate such automated vehicles in urban areas,
supervising the execution of their specified target behavior through an online monitor is key.
Due to the high level of system complexity,
novel monitoring methods are needed that allow the system to evaluate its own performance,
determine when the performance limits are reached \parencite{nolte_supporting_2020}
and when it is not capable of fulfilling the dynamic driving task anymore \parencite{sae_j3016_2021}.
This is especially important due to the absence of a human driver in case of SAE Level 4 \parencite{sae_j3016_2021}.
Research projects such as the \unicar\ and \autotech\ project address the complexity of Level 4 automation,
primarily under challenging urban conditions \parencite{woopen_unicaragil_2018, van_kempen_autotechagil_2023}.
Recent publications propose to shift the focus
from the development of robust systems with redundant processing paths
towards the development of \qemph{self-aware} road vehicles.
Self-awareness is a holistic concept that strongly relies on an explicit representation of system knowledge.
This enables the system to perceive its performance boundaries at runtime
and respond appropriately to maintain the vehicle's safe operation \parencite{nolte_supporting_2020}.
According to \citeauthor{gregory_self-aware_2016},
a key requirement for a system to become self-aware is its ability to assess its current capabilities
and project them into the future \parencite[p. 2]{gregory_self-aware_2016}.
We refer to this central task as \qemph{capability monitoring}.
By considering its capabilities,
the system can ensure that it behaves according to its specification limits.
Conversely, a decrease in performance may force the automated vehicle to adapt its behavior.

In this paper, we describe the concept and implementation
of an online capability monitor for an automated vehicle.
In addition, due to the complexity in the overall system,
we propose an expert-driven modeling approach.
The approach mainly requires two steps that are described in this work:
First, we describe the derivation of a model that is suitable to represent the relations between
the quality of the system's capabilities and its logical/technical elements.
Second, inspired by the work of \textcite{smith_bayesian_2012} and \textcite{liu_using_2015},
we consider a set of \qemph{expert rules} describing
the relations between the quality of the system elements
and operationalize them in a quantitative fashion:
Using Bayesian Networks as a foundation, we employ \qemph{Fuzzy Logic}
to derive conditional probabilities for the Bayesian Network.

We demonstrate our framework by designing a capability monitor for an example scenario.
Using a C++ library that implements the mathematical framework and that is developed alongside this work,
the monitor is instantiated for one of the \unicar\ vehicles and evaluated %
using recorded data from the real vehicle.
The software is able to process both binary error signals and continuous technical performance measures
to observe the quality state of the system elements.
By observing and propagating quality states through the Bayesian Network,
our monitor enables the vehicle to estimate whether its capabilities can be realized at a sufficient quality level.
We demonstrate our framework in this example using quality data acquired within the system at runtime.
While we note that describing the relations in the network more objectively %
requires numerous analyses for the identification of the relations in the future,
we believe that our expert-driven approach can serve as a ``template''
for the development of a more general framework.

The remainder of this paper is structured as follows:
In the following \autoref{sec:related_work}, we elaborate on related work in the field of online monitoring.
As an introduction, \autoref{sec:capabilities} first elaborates
on the concept of capabilities and the implications for
monitoring an automated vehicle using capability graphs as a basis.
We present our expert-driven framework for capability monitoring in \autoref{sec:expert},
which exploits the relationships between multiple architectural views.
After defining directed acyclic graphs (DAG)
that represent the propagation of quality states through the system
for different maneuvers, we describe the parameterization of Bayesian Networks
as a mathematical representation thereof.
In \autoref{sec:eval}, we instantiate a capability monitor for an example scenario based on our framework
and use technical measurements from one of the \unicar\
vehicles to assess the system's capabilities in a simulation.
We summarize our approach in \autoref{sec:conclusion}.
\section{Related Work} \label{sec:related_work}
In the context of automated driving,
the need for system monitoring in general and capability monitoring
in particular is widely discussed in the literature.
\textcite{maurer_flexible_2000} argues that any system component should provide quality measures that
must be aggregated by a performance monitoring function and assessed when behavioral decisions are made.
Therefore, the system requires a self-representation which,
among others, contains knowledge about its capabilities.
\textcite{reschka_fertigkeiten-_2017} describe the concept of \qemph{skill} and \qemph{ability graphs}.
According to their work, skill graphs allow to model a system by its capabilities
while ability graphs refer to the runtime implementation of skill graphs with associated performance measures.
Through the propagation of performance measures through an ability graph,
the overall system performance with respect to a considered \qemph{maneuver} can be evaluated at runtime.
Furthermore, \textcite{nolte_towards_2017} describe how skill graphs can be applied
in the design phase of an automated vehicle to identify and structure capabilities,
assign functional components that contribute to those capabilities, and derive safety requirements
that are associated with the system's functions.
Even though the concept of capability monitoring is discussed, %
to the authors' knowledge, a %
framework that explicitly incorporates the concept of capability graphs
and their architectural correspondences	for online monitoring is not publicly available yet.

In the context of health monitoring for automated vehicles,
\textcite{gomes_health_2021} present a framework that relies on \qemph{Dynamic Bayesian Networks}
to store knowledge about the relations between degradations at a component-level and overall system health.
The authors use a tree structure to represent the dependencies between the components
and a so-called \qemph{macro sub-system layer} at which multiple components are monitored.
Probabilistic graphical models such as (Dynamic) Bayesian Networks are useful to
express system interdependencies and uncertainty when monitoring the variables of a complex system.
They are therefore well-established in the literature, e.g., with respect to system diagnosis.
For example, \textcite{huang_bayesian_2008} design a
Bayesian framework for the synthesis of a control loop monitor.
\textcite{chen_applications_2011} presents the application of
Bayesian Networks in fault diagnosis for braking systems.
In a US patent, \textcite{reschka_methods_2023} outline methods for
modeling the uncertainty in different system parts through a directed graph.
As a solution, Bayesian Networks are proposed.

According to \textcite{pan_fuzzy_1998}, Bayesian Networks and Fuzzy Logic
can be combined to explicitly consider expert knowledge in the modeling process.
In particular, they should be considered as complementary concepts
representing the imprecision and uncertainty in the utilized expert knowledge.
\textcite{smith_bayesian_2012} present a probabilistic framework
for the online assessment of sensor data quality.
Using Hidden Markov Models and Fuzzy Logic, the hidden quality
state of a sensor used in marine applications can be observed.
Apart from the technical domain,
\textcite{liu_using_2015} present an expert-based system that utilizes a Bayesian Network
to assess the future population status of an animal species with respect to environmental factors in its ecosystem.
For the determination of the conditional probabilities in the network,
their work relies on expert statements represented using Fuzzy Logic.
Their work provides a valuable foundation for our approach.
\section{Capabilities in Automotive System Design}\label{sec:capabilities}
As a foundation of this work, the concept of capabilities is introduced in more detail first.
\textcite{wasson_system_2015} states that ``[a capability] represents
a physical potential -- [e.g.,] strength, capacity [or] endurance -- to perform an 
outcome-based action for a given duration under
a specified set of operating environment conditions'' \parencite[p. 229]{wasson_system_2015}
as well as ``[...] at a specified level of performance [...]'' \parencite[p. 27]{wasson_system_2015}.
A system is then defined as ``an integrated set of interoperable elements,
each with explicitly specified and bounded capabilities, working synergistically [...]
to satisfy mission-oriented operational needs in a prescribed operating environment [...]''
\parencite[p. 18]{wasson_system_2015}.
\citeauthor{wasson_system_2015} emphasizes that the description of a system's capabilities
requires both an abstract functionality and performance to be specified \parencite[p. 22]{wasson_system_2015}.
In the context of automotive system design, the concept of capabilities is,
among others, discussed in the work of \textcite{reschka_fertigkeiten-_2017}.
\citeauthor{reschka_fertigkeiten-_2017} use the term \qemph{skill} instead of capability to denote
``an activity of a technical system which has to be executed to fulfill [its] defined goals''
(\parencite[p. 3]{nolte_towards_2017}, after \parencite{reschka_fertigkeiten-_2017}).
While \citeauthor{reschka_fertigkeiten-_2017} additionally define
the term \qemph{ability} for a skill with an assigned performance measure,
\textcite{nolte_supporting_2020} note that their definition of
the terms skill and ability should be generalized under the term \qemph{capability}.

\subsection{Capability Graphs} \label{sec:arch_viewpoints}
To support the design phase of an automated vehicle,
the idea of so-called \qemph{skill graphs} is presented in the work of \textcite{reschka_fertigkeiten-_2017}.
We denote these structures as \qemph{capability graphs} hereafter.
Capability graphs are directed acyclic graphs (DAGs)
which are used as a tool to model a system by its capabilities.
A capability graph displays how particular capabilities rely on others
and how their performance affects them \parencite{bagschik_systems_2018}.
In a capability graph, the automated vehicle's capabilities
that are required to realize a specific driving maneuver \parencite{jatzkowski_zum_2021} are disassembled
into capabilities that are required for their fulfillment. %
According to \textcite{reschka_fertigkeiten-_2017}, various scenarios must be sampled from the system's specified
operational design domain and required capabilities for the target behavior of the vehicle are identified.
While this process is in this paper considered as an expert-based procedure, %
approaches towards the formal specification of target behavior exist
(e.g. \parencite{beck_phenomenon-signal_2022}). %
To cover all different parts of the vehicle's external behavior,
\textcite{reschka_fertigkeiten-_2017} proposes to construct
a set of capability graphs based on the set of possible maneuvers,
assuming that different capabilities are required for each action. %
While identifying all required capabilities requires an investigation of various scenarios,
performance requirements %
depend on the specific task, i.e., are scenario-dependent \parencite{bagschik_systems_2018}.

\subsection{Architectural Considerations} \label{sec:arch}
Capability views are considered prior to the specification of functional views,
which display the information flow through the system.
In the architectural considerations made by \textcite{bagschik_systems_2018},
correspondences between capability, functional, software and hardware views are described.
For each capability of the system, requirements are considered to constrain the design of specific functions,
which contribute to the realization of the respective capability \parencite{nolte_towards_2017}.
Consistent with the ISO 26262 life cycle, %
components of the functional architecture are assigned to the functional requirements.
In the subsequent step, hardware and software (HW/SW) architectures are designed,
which allow subsequently defined technical requirements
to be assigned to components \parencite{bagschik_systems_2018}.
In the \unicar\ project, the \qemph{Automotive Service-Oriented Architecture} (ASOA)
\parencite{kampmann_dynamic_2019} is developed and applied:
Software services are introduced to implement the elements of the functional architecture and
hardware components are then selected for the services to be executed on. %
As an intermediate layer between the functional and HW/SW views,
a logical view is considered to allow abstracting from technological solutions
for the functions to be implemented \parencite{walden_incose_2015}.
\begin{figure}[H]
    \vspace{-0.35em}
    \centering
    \includegraphics[width=0.95\columnwidth, height=4.00cm]{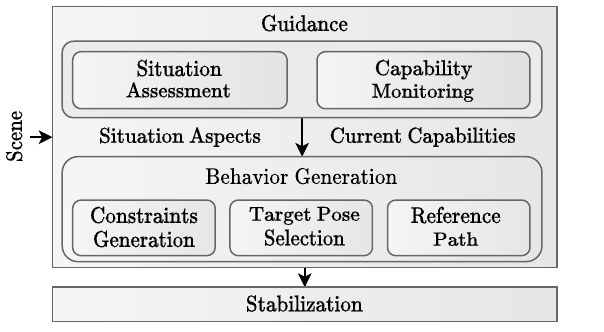}
    \vspace{-1.00em}
    \caption{%
        Functional architecture extract, adapted from \parencite{nolte_model_2017, ulbrich_towards_2017}.%
    }
    \label{fig:funcarch}
    \vspace{-0.35em}
\end{figure}
\subsection{Capability Monitor} \label{sec:monitor}
A decline in system health at runtime
(e.g., due to degradations or failures of hardware components)
as well as performance insufficiencies are expected to inhibit the system
from realizing the required capabilities and hence the specified behavior.
Such declines therefore need to be detected and indicated to the decision making module.
Accordingly, the automated vehicle should only perform maneuvers
for which it has the required capabilities.
This has important implications for the interaction between
monitoring and runtime decision making of an automated vehicle:
\textcite{nolte_model_2017} consider a system-level monitor
(``\textit{Ability Monitoring}'' \parencite[Fig. 2]{nolte_model_2017})
as an input for the decision making module at the guidance layer
of their functional architecture.

In this work, we employ a \emph{capability monitor}
as a system-level monitor that is able to indicate the quality of the system's capabilities at runtime.
In terms of the functional architecture,
we establish our monitor as a direct input to the behavior generation (see \autoref{fig:funcarch})
as in \parencite{nolte_model_2017}.
In terms of the HW/SW architectures, we realize the capability monitor
as an ASOA service executed on the $\qemph{StemBrain}$ ECU \parencite{keilhoff_unicaragil_2019}
of the \unicar\ vehicle.
This is, we employ a single centralized monitor %
that relies on the quality information of various distributed system elements.
Further decentralizing the monitoring task is possible but not discussed.
\section{Expert-Driven Framework} \label{sec:expert}
The general description raises the question of which
technical variables need to be monitored exactly and how they interact.
\textcite{nolte_towards_2017} argues that the selection of technical variables for monitoring
should be based on capability-level requirements
that are formulated with respect to the system's desired behavior.
Such behavioral requirements address, among others, quantities related to
the vehicle's (externally observable) motion, e.g., the required minimum lateral
deviation from a reference (as in \parencite{nolte_towards_2017})
or maximum deceleration (as in \parencite{bagschik_systems_2018}).
Technical determinants -- e.g., the available braking force
or the accuracy of the vehicle's state estimation --
hence influence the quality of the capabilities and are therefore crucial to be monitored.

In order to model the interdependencies between capability-level performance measures
and, e.g., contributing elements at the HW/SW level, numerous quantitative analyses 
should be conducted to derive models,
which then allow using monitored variables to infer the quality of the system's capabilities.
However, conducting numerous analyses requires time and resources.
In an early stage of the development process, a practical first step %
would be to consult domain experts that help to identify relevant factors and relationships:
In line with \parencite{liu_using_2015},
we claim that expert judgement is often required (a) to identify performance determinants,
(b) select and conduct meaningful analyses, (c) interpret the results and
(d) finally execute the modeling process to capture the relationships.
This is particularly true in (the early stages of) an \qemph{agile} development process
where the system architecture may be subject to frequent changes.
Our experience from the \unicar\ project confirms this intuition.

In this paper, we focus on the early stages of the development process
and therefore note the usefulness of expert knowledge
in order to generate a template for a system model used for monitoring.
However, basing the supervision of a safety-critical system on the intuition of experts
would be rather critical when actually releasing such a system.
Nonetheless, we claim that our work points out some challenges to be addressed in future work:
the overwhelming complexity and uncertainty in dealing with automated driving systems
and the inherent need for expert input.

\subsection{Selected Requirements} \label{sec:requirements}
Hereafter, we present the steps in our framework to design a capability monitor for an example scenario.
Of course, this framework shall follow certain requirements.
Given \autoref{sec:capabilities}, we state that
a system monitor should be designed using the capability concept as a base.
We therefore need the framework to explicitly rely on the capability graphs
derived in an early stage of the design process. %
Given the capabilities' importance for the system's behavior,
we further require a monitor derived from our framework to be able to inform the system's decision making,
i.e., at least allow the selection of an admissible set of maneuvers given a certain context. %

Following the examples in \parencite{nolte_towards_2017, reschka_fertigkeiten-_2017}, 
we also aim to derive an augmented system model for monitoring purposes,
which extends the capability graph representation by including implementation-specific elements
(e.g., HW/SW components) to relate their current quality state to the quality of the capabilities.
We propose to use a DAG representation.
We then require suitable mathematical representations %
for the propagation of quality information through the graph, even under uncertainty.
In the following, we use Bayesian Networks, but other methods could be used.

Finally, to close the gap, we need the framework to allow
for the operationalization of expert knowledge in a systematical way,
which can be interpreted in a quantitative and/or logical fashion.
This enables the automatic inference of system performance at runtime.
Note that this requirement becomes abundant if a sufficient amount of analyses
has been conducted to describe every single relationship in the system model objectively.

\subsection{Example Architecture Description} \label{sec:model}
Different architectural views of the system allow addressing different aspects.
While the capability view yields the highest level of abstraction and is directly related
to the system's behavior, the HW/SW and logical views allow to address
the low-level determinants of the system's capabilities that can be measured instantaneously at runtime.
Exploiting the correspondences between these views (see \autoref{fig:arch_viewpoints})
can therefore help to trace system interdependencies from abstract capabilities to technical variables.
Therefore, the first step in our framework is to consider the system's architecture using multiple views
and to trace the relationships between the quality of system elements across them.
As an example, we give a brief informal description of selected capabilities of the \unicar\ vehicles
as well as the correspondences with respect to other architectural views.
In the following section, we will use this example to derive the desired quality propagation model accordingly.

\paragraph*{Capability View}
The capability view is adapted through capability graphs.
As an example, we only focus on the \unicar\ vehicle's longitudinal motion:
Among others, the vehicle shall have the capability of controlling
its longitudinal motion in order to follow a desired speed.
In the capability graph, the vehicle's capability of ``longitudinal motion control''
can be further decomposed into the capabilities ``accelerate'', ``decelerate'' and ``estimate motion''.

\paragraph*{Functional View}
A trajectory tracking controller \parencite{homolla_encapsulated_2022}
is introduced in the functional architecture that contributes to the vehicle's control capability.
Furthermore, powertrain and brake actuation functionalities are required
contributing to the capabilities ``accelerate'' and ``decelerate''.
A functional element for (vehicle) dynamic state estimation (VDSE) \parencite{gottschalg_integrity_2020}
is introduced contributing to the capability ``estimate motion''.

\paragraph*{Logical View} See also \parencite{keilhoff_unicaragil_2019}.
The controller is implemented for the motion control function
through a control algorithm that is executed on the so-called ``StemBrain'' ECU.
For the powertrain, four wheel-individual electric motors and
corresponding power electronics are implemented in the \unicar\ vehicle.
Since the powertrain is able to provide both positive and negative torque,
it is able to accelerate and decelerate the vehicle.
The brake functionality is implemented in hardware by a two-unit hydraulic brake system with a hydraulic power unit.
For both the powertrain and brake components,
we summarize the motors/brake units and correspondings power electronics as logical elements.
The so-called ``SpinalCord'' ECU provides the required communication for both \parencite{keilhoff_unicaragil_2019}.
Finally, the state estimation function utilizes an IMU, GNSS receiver and odometry sensors,
and a \emph{localization filter} to fuse their signals --
which we consider as another logical element \parencite{gottschalg_integrity_2020}.
\subsection{Deriving the Graph Representation} \label{sec:quality}
\begin{figure}[H]
    \vspace{-0.55em}
    \centering
    \includegraphics[page=7, width=1.0\columnwidth, height=4.35cm]{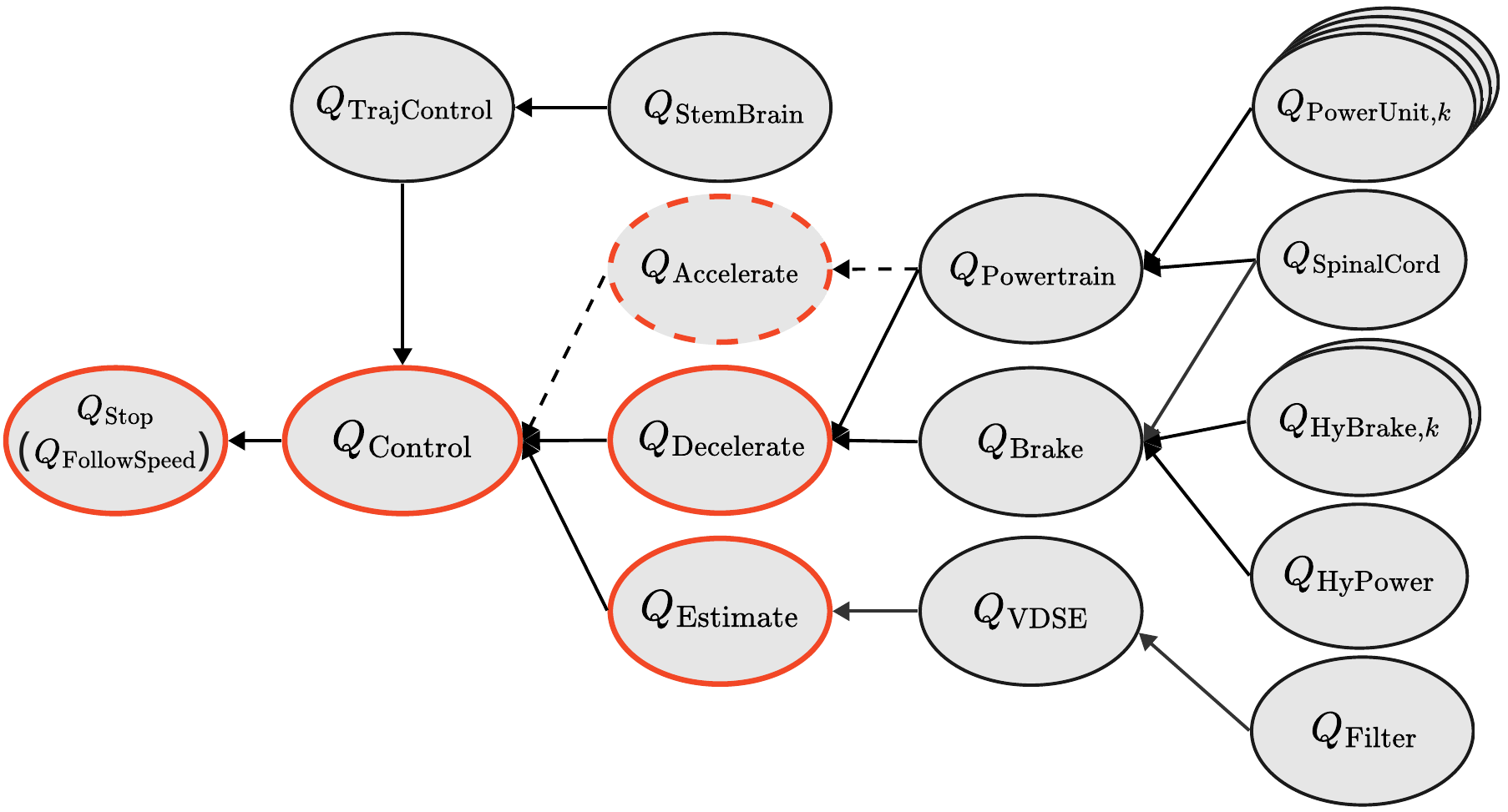}
    \vspace{-1.15em}
    \caption{%
        A network displaying the interdependencies between the quality of elements
        across different architectural views for the longitudinal \qemph{stop} and \qemph{follow speed} maneuver.
        The latter requires additional capabilities, i.e., nodes and edges indicated by dashed lines.%
    }
    \label{fig:capcomp}
    \vspace{-0.75em}
\end{figure}

Although the elements of different architectural views have a different semantic,
we claim that they are comparable in terms of their quality.
I.e., we suggest that the abstract quality of a capability can be mapped
to the quality of elements in the functional and logical view.
Accordingly, we define a system model that incorporates these cross-architectural relations.
We define it as a DAG where the nodes represent the \emph{quality states} $Q_i$
of the system elements and the edges represent correspondences between them.
Hereafter, the derivation of the DAG is a hand-tuned expert-based procedure.
The resulting model is shown in \autoref{fig:capcomp} for the \qemph{stop} and \qemph{follow speed} maneuvers.
Following \parencite{reschka_fertigkeiten-_2017},
we note that a DAG definition for each maneuver would be required
as each of them relies on different capability sets (graphs).
We continue our previous example of the vehicle's longitudinal motion control.
As mentioned before, in order to control its motion,
the automated vehicle must also have the capabilities ``accelerate'', ``decelerate'' and ``estimate motion''
for which we introduce the quality states
$Q_{\mathrm{Control}}$, $Q_{\mathrm{Accelerate}}$, $Q_{\mathrm{Decelerate}}$ and $Q_{\mathrm{Estimate}}$.
Since several functions are contributing to the realization of these capabilities
(motion controller, powertrain/brake actuation and state estimation functions),
we further introduce $Q_{\mathrm{TrajControl}}$, $Q_{\mathrm{Powertrain}}$,
$Q_{\mathrm{Brake}}$ and $Q_{\mathrm{VDSE}}$.
The controller is implemented on the ``StemBrain'' ECU with $Q_{\mathrm{StemBrain}}$.
Furthermore, the four wheel-individual electric motors and corresponding power electronics
(\qemph{powertrain units}) yield the quality states $Q_{\mathrm{PowerUnit},k}$.
Their performance is directly influencing the performance of the powertrain.
In terms of the braking functionality,
we further denote the performance of each brake unit as $Q_{\mathrm{HyBrake},k}$
and the state of the hydraulic power unit as $Q_{\mathrm{HyPower}}$.
Both are directly linked to the quality state $Q_{\mathrm{Brake}}$ in the DAG.
The powertrain and brake components rely on the communication that is
provided on the so-called ``SpinalCord'' ECU with state $Q_{\mathrm{SpinalCord}}$.
Finally, we consider the state estimation function and
capture the quality of the implemented filter in $Q_{\mathrm{Filter}}$.
\subsection{Design of a Bayesian Network} \label{sec:math}
Given the DAG structure,
we aim to establish a mathematical model representing the interaction between the quality states.
We expect purely deterministic approaches to be unsuitable due
to the high level of uncertainty in the modelling process and
therefore propose the use of \qemph{Bayesian Networks}.
Dynamic models incorporating time-dependent effects -- e.g., as in \parencite{gomes_health_2021}
-- may be considered in future work.
Based on that, we do not assign a particular quality state to each node
but a distribution of \qemph{beliefs} (of being in a specific state).
In a Bayesian Network, the belief in the state of
any node is assumed to only rely on the $N$ parent nodes,
$\Pri{\Qi}{\mathbf{Q}_{j}} = \Pri{\Qi}{Q_{j_1},\dots, Q_{j_N}}$.
Hence, the state of any node can be derived by evaluating these conditional probabilities along the graph.
At the bottom of the graph in \autoref{fig:capcomp}, nodes without any parents exist.
We refer to these nodes as \qemph{input nodes} of the network.
The quality of these nodes cannot be infered but must be observed based on additional measures
(see \autoref{sec:measures}).
If the set of conditional probabilities as well as additional observations are available,
the state of any node in the network can be inferred.
For a two-step example, we obtain
\begin{align} \label{eq:infer}
    \begin{split}
    & \Pri{\Qi,\,\Qjo,\dots, \Qjn}{\mjoout,\dots ,\mjnout} \\
    = & \Pri{\Qi}{\Qj} \cdot \Pri{\Qjo}{\mjoout}\cdot\dots\cdot\Pri{\Qjn}{\mjnout}.
    \end{split}
\end{align}
for the combination of states $\Qi,\,\Qjo,\dots, \Qjn$.
As we design our framework to be used by experts, we follow \parencite{smith_bayesian_2012}
and simplify the state space to %
\begin{equation} \label{eq:states}
    \begin{split}
        \Qi \in \Q = %
        \{ S_1 =\good,\; S_2=\pgood, \\
        S_3 =\pbad,\; S_4=\bad\}%
    \end{split}
\end{equation}
Of course, this representation neglects actual physical properties --
however, we assume that it is the only form that experts can handle intuitively.
The distribution of beliefs for each node $i$ is then expressed as the normalized vector
\begin{equation} \label{eq:vpi}
    \vpi = \left [ \Pri{\Qi = S_1}{\mathbf{Q}_{j}} ,\dots, \Pri{\Qi = S_4}{\mathbf{Q}_{j}} \right ]^T.
\end{equation}
As shown in \eqref{eq:infer}, the vector of beliefs $\vpi$
depends on the belief distributions for the states of each parent node
as well as the conditional probabilities describing the relations between them.
In a Bayesian Network with such discrete states,
the conditional probabilities are expressed through conditional probability tables (CPTs).
In order to incorporate expert statements,
we follow the work of \textcite{liu_using_2015} to generate these tables:
For each node in the network, we introduce a set of logical
if-then rules (\qemph{Fuzzy Rules}) that determine
the outcome of every possible state combination of the inputs.
Every combination of parent states (antecedents)
must be assessed by experts and a resulting state (consequent) of the child node is assigned. %
Each statement is stored as a rule $\srule$ with the form
\begin{quote}
    \textit{If} $Q_{j_1} = S_{j_1}$ $\land$ $\dots$ $\land$ $Q_{j_N} = S_{j_N}$ \textit{then} $Q_{i} = S_{i}$,%
\end{quote}
\vspace{-0.15em}
\noindent %
e.g.,
\begin{quote}
    \textit{If} $Q_{\mathrm{Estimate}} = \good$ $\land$ $Q_{\mathrm{Decelerate}} = \pdgood$  $\land$ $Q_{\mathrm{Accelerate}} = \bad$ \textit{then} $Q_{\mathrm{Control}} = \bad$.%
\end{quote}
\vspace{-0.15em}
\noindent %
In general, despite the small number of four possible states per node,
the overall number of expert rules is large,
which allows representing complex relationships
that are still traceable when viewed and compared in detail.
To let a machine interpret such rules,
we apply the \qemph{Mamdani implication} \parencite{mamdani_experiment_1975}
with the max-product operator as in \parencite{liu_using_2015}. %
It allows to consider the set of expert rules and automatically determine the set of CPTs for the Bayesian Network.
When using the Mamdani implication, the state of a child node is not necessarily
directly dependent on the consequent for the combination of antecedents
specified in the corresponding expert rule: %
So-called \qemph{membership functions} are defined for each node and its states,
which are applied to the child nodes' states (antecedents of the rule).
If the defined membership functions of the antecedents overlap, an interaction between the child nodes' states occur,
distributing the belief over multiple states of the considered node (see \parencite[Fig. 4]{liu_using_2015}).
This can be used to model the uncertainty in the system.

In order to apply the Mamdani implication,
each (parent) node requires a membership function for each possible state to be set, $\mustatevnn$.
Note that $\pidx_n$ for $n\in 1, ... ,N$ denotes the considered parent node
while $i$ refers to the child node.
The conditional probability is then defined as follows:
\begin{align} \label{eq:mamdani}
    \begin{split}
    & \Pri{\Qi = \statevi}{\Qjo = \statevo,\dots, \Qjn = \statevn} \\
    & = \max_{\srule\in\srulesv} \left \{
            \prod_{n=1}^N \mustateann\left ( \arg\max\left \{ \mustatevnn \right \} \right ) %
        \right \}.
    \end{split}
\end{align}
with $v = 1,...,4$. %
The subset $\srulesv\subset\srules$ contains all rules $\srule$ with the consequent $\con(\srule)\equiv \statevi$.
The membership function of the $n$-th antecedent of the rule $\srule$ is $\mustateann$. %
In this work, only Gaussian membership functions are used for this purpose.
The expression to the right of the product sign in \eqref{eq:mamdani}
can be seen as a measure of ``compatibility'' between the currently evaluated rule $\srule$,
the antecedents addressed in it and the currently considered states of the parent nodes.
Only if membership functions overlap, the belief is distributed over multiple states of the child node.
During the inference process, the four strings in $\mathcal{Q}$
are internally represented through the values $S_1=0,\dots,S_4=3$
and each Gaussian membership function is centered around the respective value.

\subsection{Continuous Belief Transformation}
Since the quality information for each node $i$ is encoded in four belief values,
the normalized vector in \eqref{eq:vpi} must be evaluated to interpret the state of any node.
However, we note that the comparison of two vectors in general
-- i.e. two nodes inside the same network or the same node under different conditions --
is not directly possible and hence, the interpretation of the inferred values in each node is difficult.
We propose to aggregate the data into a single scalar quality indicator --
referred to as \qemph{continuous belief} value:
\begin{equation} \label{eq:cbelief}
    \cbeh := 0.5 + (\vpi^{(1)} + w \vpi^{(2)} - w \vpi^{(3)} - \vpi^{(4)})/2\in [0,\, 1].
\end{equation}
The transformation incorporates the symmetric weighting of the belief values
with respect to the \qemph{(probably) good} and \qemph{(probably) bad} state
as well as a weighting between the \qemph{good/bad} and \qemph{probably good/bad}
state that can be varied via the single parameter $w$.
The continuous belief could hence be seen as the ``percentage'' of belief in a \qemph{good} quality.
\subsection{Selection of Technical Measures} \label{sec:measures}
\begin{figure}
    \centering
    \includegraphics[width=0.96\columnwidth]{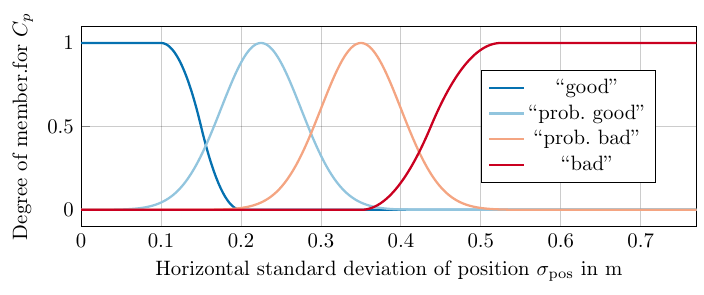}
    \caption{Membership functions for the horizontal standard deviation of position
        as a performance measure for the state estimation function.}
    \label{fig:mufzs}
\end{figure}
The problem of capability monitoring is in this work reduced to the evaluation of the nodes in a Bayesian Network.
There, additional measurements are required to make estimates for the input nodes.
Such measures can be usually separated into continuous measures -- e.g., expressing signal levels -- %
and binary states such as availability or error indicators. %
The selection of feasible technical measures for this purpose incorporates design choices made by experts.
In mathematical terms, the measure $\miout$ is an observation that allows to estimate the state of an input node.
Following our Fuzzy Logic based approach, we apply the membership function $\muobs$ to evaluate the value of $\miout$.
The function returns the degree of membership for the observed measure with respect to the \qemph{Fuzzy Set} for the state $\Qi$,
which can be equated with the belief for the node to be in this state,
\begin{equation} \label{eq:observe}
    \Pri{\Qi = \statevi}{\miout} = \muobs(\miout),\quad v = 1,...,4.
\end{equation}
To demonstrate our approach, we consider observations for the Bayesian Network that are based on the model displayed in \autoref{fig:capcomp}.
First, we consider the vehicle's capability ``estimate motion'' to which the state estimation function contributes to:
Based on the applied localization filter,
we consider the standard deviation in the filtered signal for the vehicle's horizontal position to be a meaningful quality measure. %
We prefer an indicator based on the global position since it integrates deviations in the acceleration and velocity.
For the derived quality measure, appropriate membership functions are defined to evaluate observations as shown in \autoref{fig:mufzs}.
The output of the membership function determines the state of the node $Q_{\mathrm{Filter}}$.
Furthermore, in the actual implementation of the filter involving hardware components,
additional binary error states for required hardware components (e.g., ECUs) are available.
If an error flag is raised at runtime, we directly set the node's state to $\bad$ and neglect other observations.

Similarly, we consider the logical components
that contribute to the vehicle's capabilities ``accelerate'' and ``decelerate'':
The four powertrain units ($Q_{\mathrm{PowerUnit},k}$) of the \unicar\ vehicle rely on electric motors.
A decrease in the supplied electric energy as well as a degradation of the underlying power electronics,
e.g., a de-rating of the DC/AC inverter, is expected to affect the available maximum power.
For example, a decline in the battery voltage level will shift each motors' corner speed
in such a way that less torque is available along the entire torque-speed curve of the motor.
We consider the voltage level of the power electronics as a feasible technical measure
to make statements about the overall powertrain performance.
We also incorporate binary error flags which, e.g., denote the health of
the power electronics in terms of over/under voltage, current or temperature issues.

As a simplification in the following examples,
we assume the performance of the hydraulic brakes to be flawless and set their state to $\good$.
We assume the same for all ECUs. %
\subsection{Software Implementation} \label{sec:implementation}
The described mathematical framework is implemented in a C++ software library as part of this work.
During the design phase, a JSON-style \qemph{network file} specifying the network nodes,
their parent-child relations as well as associated membership functions can be defined for each DAG. %
The Fuzzy Rules are stored in a \qemph{rule file}.
Our C++ software library provides the code for automatically
inferring the CPTs using the Mamdani implication offline.
At runtime, the inferred CPTs are used to determine the state of each node in the network
while continuously observing measures $\miout$ at the input nodes.
For the online inference process, the \qemph{libDAI} library is used \parencite{mooij_libdai_2010},
which also allows to directly set fixed values for the states online, e.g., to handle binary error flags.
\section{Application in an example scenario} \label{sec:eval}
\begin{figure}[H]
    \centering
    \includegraphics[page=1, width=0.97\columnwidth]{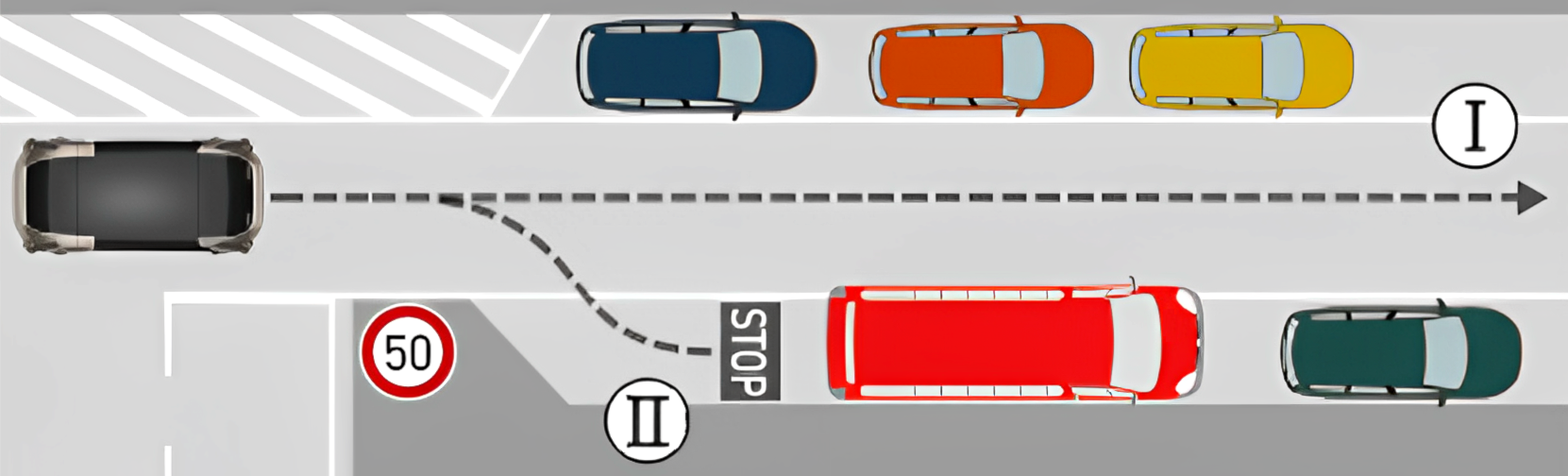}
    \vspace{-0.15em}
    \caption{
        Example scenario -- adapted from \parencite{stolte_towards_2020}. Not true to scale.
    }
    \label{fig:scene}
\end{figure}

In this section, we demonstrate the application of the described framework
and deploy an online capability monitor in an example scenario.
The scenario is similar to an example in \parencite{stolte_towards_2020}.
We refer to the \unicar\ vehicle in this scenario as the ego-vehicle.
The start scene is visualized in \autoref{fig:scene}.
The ego-vehicle on the left is driving on a straight one-way road segment at the speed limit of $50\tkmh$.
Here, we assume that ego-vehicle may perform the \qemph{follow speed} or \qemph{stop} maneuver.
The expected paths resulting from the vehicle's motion are visualized
as options I (\qemph{follow speed}) and II (\qemph{stop}).%

In the following, we neglect the lateral motion %
and only focus on the two different longitudinal maneuver options.
Therefore, our capability monitor should contain the DAG representations for these two options
as discussed in subsections~\ref{sec:model} and \ref{sec:quality},
This is, we use the DAG representations and store them in network files. %
For the sake of brevity, the implementation of the example
is limited to the system parts shown in \autoref{fig:capcomp}.
For the scenario and the two maneuvers, an expert survey is conducted, and expert rules are stored.
Finally, we deploy the inference engine together with the network and rule files,
and evaluate its behavior in a simulation.
\subsection{Example Data and Events} \label{ref:scenario}
\begin{figure*}[!t]
    \centering
    \includegraphics[width=1.0\textwidth]{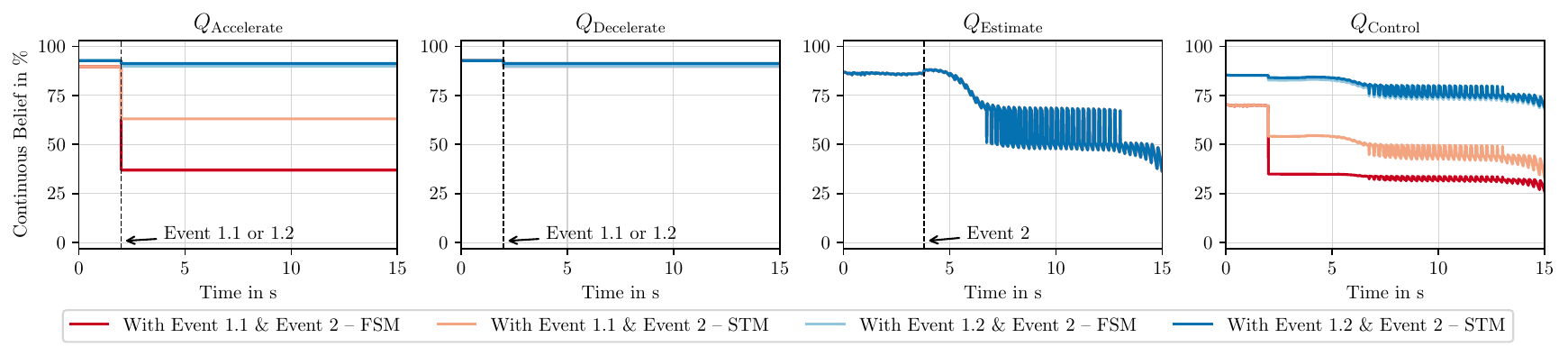}
    \vspace{-2.0em}
    \caption{Continuous belief for the quality state of nodes in the Bayesian Networks for
        the \qemph{follow speed} ($\mathrm{FSM}$) and \qemph{stop} maneuvers ($\mathrm{STM}$).}
    \label{fig:scenario1}
    \vspace{-1.5em}
\end{figure*}
In the simulation
we use recorded data streams acquired online within the ASOA service framework.
The sample time of the simulation is $T_S = 0.01\s$. %
As we aim to demonstrate the monitor's output in case of degradations or performance insufficiencies appearing at runtime,
we simulate their influence by manipulating the recorded data where necessary.
In the simulation, we consider the following possible events:
\begin{itemize}
    \itemsep0.1em
    \item \textbf{Event 1.1}:
    After $2\s$, an isolated E/E failure in a single electric motor is assumed to appear and to be detected instantaneously.
    The failure has no effect on the other motors. %
    The error signal sets the quality of the corresponding node in the networks in \autoref{fig:capcomp} %
    to $\bad$ while the quality of the remaining units is $\good$.
    \item \textbf{Event 1.2}:
    In an alternative version of the first event, E/E failures in two electric motors are assumed to appear after $2\s$.
    It is handled as in Event 1.1.
    \item \textbf{Event 2}:
    After $3.8\s$, an increase in the standard deviation measured in the localization filter is detected.
    One possible reason for this performance insufficiency is the shadowing of the GNSS sensor by nearby buildings. %
\end{itemize}
\subsection{Parameterization} \label{sec:eval_parameters}
In each time step of the simulation, %
the quality state of each node in each of the two Bayesian Networks is inferred.
To aggregate the measurements, we consider the continuous belief $\cbeh$ in the following.
Even though our framework relies on qualitative expert knowledge,
quantitative quality states are produced for quantitative data at the input nodes.
In the following, we set $w=0.33$ in \eqref{eq:cbelief}.
For all Gaussian membership functions $\mustateann$, %
we set a constant standard deviation of $0.3$ in \eqref{eq:mamdani}, see \autoref{sec:math}.
The underlying expert rules are available online.\footnote{%
    \url{https://git.rz.tu-bs.de/r.schubert/rule_files}%
}

\subsection{Results} \label{sec:results}
Given \autoref{fig:capcomp}, we focus on
$Q_{\mathrm{Control}}$, $Q_{\mathrm{Accelerate}}$, $Q_{\mathrm{Decelerate}}$ and $Q_{\mathrm{Estimate}}$
and their continuous belief values $\cbeh$.
Note that the capability ``accelerate'' is not required in case of the \qemph{stop} maneuver.
Since we only focus on an extract of the system and a longitudinal maneuver,
the quality state of the vehicle's capability of performing the respective maneuver
is only influenced by the vehicle's capability ``longitudinal control''.
Hence, we reduce our assessment to the capability ``longitudinal control''.
In the DAG, $Q_{\mathrm{Accelerate}}$ solely relies on $Q_{\mathrm{Powertrain}}$.
Due to control allocation techniques applied in the controller \parencite{homolla_encapsulated_2022},
a certain degree of robustness is introduced into the system. %
In the rule set, the powertrain's quality is still rated (by experts)
as $\pgood$ with respect to the availability of three out of four powertrain units
-- given the current urban scenario and speed limit.
In the simulation, at the moment of the partial powertrain failure (Event 1.1),
the value of $\cbehnoidx_{\mathrm{Accelerate}}$ declines from $90\perc$
and reaches a value of approximately $63\perc$.
Due to the availability of the hydraulic brake system, $\cbehnoidx_{\mathrm{Decelerate}}$
is only slightly affected.

In case of Event 1.2, due to the lack of power available for realizing the capability ``accelerate'',
the powertrain's quality is rated (by experts) as $\pbad$ with respect to the availability
of two out of four powertrain units.
$\cbehnoidx_{\mathrm{Accelerate}}$ reaches a value of approximately $37\perc$.
The capability ``decelerate'' reaches a minimum continuous belief value of $90\perc$ in this case.
For the quality of the capability ``estimate motion'', the results are unchanged until
this point in time but is affected due to Event 2:
Odometry-based sensing is not active in the recorded dataset, thus the shadowing
of the GNSS signal results in a wide noise band.
This effect is also visible in the continuous belief value for the capability ``estimate motion''.
In contrast to the constant results
for the other two capabilities, $Q_{\mathrm{Estimate}}$ changes continuously.
$\cbehnoidx_{\mathrm{Estimate}}$ reaches a minimum of approximately $37\perc$.

In the last step, we focus on $Q_{\mathrm{Control}}$ %
for the four combinations of the two longitudinal maneuvers
(\qemph{follow speed} or \qemph{stop}) and two events (Event 1.1 or Event 1.2 and Event 2).
In case of the \qemph{follow speed} maneuver, a worst-state assumption is made by experts
regarding the quality of the required capabilities,
i.e., the quality of performing the maneuver is set equal to the lowest quality of
any required capability in the rule set.
The degradation introduced through Event 1.1 results into a value of
$\cbehnoidx_{\mathrm{Control}} \approx 54\perc$. %
In case of Event 1.2, this value is below $34\perc$ even before Event 2.
In both cases, the low quality of the capability ``accelerate''
is the dominant factor implicit in the expert rule base.
After Event 2, the value of $\cbehnoidx_{\mathrm{Control}}$ decreases
with a minimum of $34\perc$ after Event 1.1.
In case of Event 1.2 and Event 2 happening, a minimum of approximately $27\perc$ is reached.

For the \qemph{stop} maneuver, the quality of the vehicle's ability to ``decelerate''
is considered crucial by experts
and accordingly weighted with large weight in the rule set.
In case of the \qemph{stop} maneuver, the resulting continuous belief
is only slightly affected by the partial failure of the powertrain in Event 1.1 and 1.2
as $\cbehnoidx_{\mathrm{Decelerate}}$ only shows a minor decline as well.
For both combinations of Event 1.1/1.2 and Event 2,
the values of $\cbehnoidx$ for ``longitudinal control''
are much higher than for the \qemph{follow speed} maneuver.

Finally, with respect to the role of capabilities for the realization of the vehicle's behavior,
we propose to use the inferred quality information
for the capability nodes to support the runtime decision making process.
We note that a continuous belief of $50\perc$ marks the symmetrical boundary
between the $\good$ and $\bad$ quality range
as the quality states are equidistantly distributed (see \autoref{sec:math})
and the continuous belief function in \eqref{eq:cbelief} employs symmetric weighting.
Hence, one could argue that a maneuver with a value of $\cbehnoidx < 50\perc$ should not be executed.
Here, with respect to the continuous belief value for ``longitudinal control'',
the maneuver \qemph{follow speed} should be eliminated from the space
of admissible actions after both events occurred.
The admissible action space should in these cases be reduced to the \qemph{stop} maneuver.
In other words, the ego-vehicle should come to a full stop as its capabilities
are not sufficient for performing another action.
\section{Conclusion and Future Work} \label{sec:conclusion}
In this paper, we present an expert-based approach towards online capability monitoring for automated vehicles.
In order to implement a monitor for the automated \unicar\ vehicle,
we propose the derivation of a DAG that represents the propagation of quality states through the system.
By operationalizing the DAG as a Bayesian Network,
we are able to infer the quality of the system's capabilities at runtime.
In our framework, we use expert knowledge in two ways.
First, for structuring the DAG model for quality propagation
for which we consult the various architectural views.
Second, we use expert knowledge to parameterize Bayesian Networks using expert rules.
In terms of the first aspect, in the future, we aim to conduct further research
with respect to the introduction of more formal architecture descriptions
that allow the derivation of such a system model in a more concise and traceable manner.
Formal modeling languages might be suitable for this purpose.

In terms of the second aspect, while acknowledging the abstraction and simplification
that the Fuzzy Logic approach introduces, we also note the loss of traceability and objectivity.
As pointed our before, we see the need for replacing the expert-driven models with models
that are derived from quantitative analyses (e.g., by using statistical methods).
This allows to preserve the physical meaning of the input-output relations between nodes.
Nonetheless, the graphical and probabilistic model structure presented in this paper is promising
as it also allows the expert-based probability distributions to be replaced by distributions
that are from such analyses (for a more complex state space).

A third, more general aspect is that our approach only addresses intro-spective self-perception
paired with performance criteria that are defined for a single scenario.
This is, our rule set is only valid for the discussed scenario.
The property of \qemph{situational awareness} is a second,
equally important aspect of self-awareness \parencite{gregory_self-aware_2016}.
Applying our framework to a large number of scenarios would not be feasible in its current form
as it would require the derivation of an extremely large number of expert rules which is impractical to obtain.
The application within a specific use case (that comprises a set of scenarios)
might be feasible, however, at least for prototypical and test purposes.
For introducing situational awareness, physical properties related to, e.g.,
the current environment and vehicle state need to be considered as inputs of the model as well.
We hope that by using more advanced and objective methods for the parameterization of the Bayesian Network,
a model that works across a wider range of scenarios can be derived in the future.
\section*{Acknowledgement}
\label{sec:acknowledgement}
We would like to thank Tobias Homolla, Stefan Leinen, Grischa Gottschalg, Hendrik Marks
and Marc Leuffen for providing their valuable expert knowledge.
Furthermore, we thank our colleagues Nayel Fabian Salem and Niklas Braun for the fruitful discussions.

\printbibliography

\end{document}